\documentclass[a4paper,12pt]{article}
\usepackage{graphicx}
\usepackage{cite}
\usepackage{amssymb,amsmath,latexsym,enumerate}
\usepackage{color}
\usepackage{bigints}

\begin{document}

\title{Analytical solutions for problems of bubble dynamics}

\author{Nikolai A. Kudryashov, Dmitry I. Sinelshchikov,\\
\texttt{nakudr@gmail.com}, \texttt{disine@gmail.com}, \vspace{0.5cm}\\
Department of Applied Mathematics, \\ National Research Nuclear University MEPhI, \\ 31 Kashirskoe Shosse, 115409 Moscow, Russian Federation}

\date{}

\maketitle

\begin{abstract}
Recently, an asymptotic solution of the Rayleigh equation for an empty bubble in $N$ dimensions has been obtained. Here we give the closed--from general analytical solution of this equation. We also find the general solution of the Rayleigh equation in $N$ dimensions for the case of a gas--filled hyperspherical bubble. In addition, we include a surface tension into consideration.
\end{abstract}



\section{Introduction}

Nowadays the Rayleigh equation for bubble dynamics and its generalizations are intensively studied. These equations are used in various applications like technology and medicine \cite{Brennen2013,Doinikov2009,Doinikov2013,Bogoyavlenskiy2000,Kudryashov2010a,Kudryashov2013}.

Recently, particular attention has been paid to the analytical studying of the Rayleigh equation. For instance, accurate asymptotic approximations for solutions of the Rayleigh equation have been obtained in \cite{Obreschkow2012,Fernandez2013}. Then, these results have been generalized to include the $N$--dimensional case of the Rayleigh equation \cite{Klotz2013}. Although it has been stated that there are no analytical solutions of the Rayleigh equation (see, e.g. \cite{Brennen2013,Obreschkow2012,Alehossein2007}), such solutions have been found in \cite{Kudryashov2014}. However, closed--form analytical solutions of the Rayleigh equation in $N$ dimensions have not been obtained previously.

In this work we construct the general closed--form analytical solutions of the Rayleigh equation both for an empty and gas--filled hyperspherical bubble. To this end we use an approach that has been proposed in \cite{Kudryashov2014}. We also take into consideration a surface tension.

The rest of this work is organized as follows. In Section 2 we give basic equations and describe our approach. Section 3 is devoted to the consideration of an empty hyperspherical bubble. In Section 4 we study the case of a gas--filled hyperspherical bubble. We consider the influence of a surface tension in Section 5. In the last section we briefly discuss our results.

\section{Main equations and approach}
Let us consider the Rayleigh equation for an empty hyper--spherical cavity in $N$ dimensions \cite{Klotz2013}
\begin{equation}
\frac{\rho}{N-2}\left(RR_{TT}+\frac{N}{2}R_{T}^{2}\right)=-p,
\label{eq:Rayleigh_ec}
\end{equation}
and the Rayleigh equation for a gas--filled hyper--spherical cavity \cite{Klotz2013}
\begin{equation}
\frac{\rho}{N-2}\left(RR_{TT}+\frac{N}{2}R_{T}^{2}\right)=
\left(P_{0}+\frac{(N-1)\sigma}{R_{0}}\right)\left(\frac{R_{0}}{R}\right)^{N\kappa}-p-\frac{(N-1)\sigma}{R}.
\label{eq:Rayleigh_nec}
\end{equation}
Here $R$ is the radius of the bubble, $T$ is the time, $p$ is the far--field pressure, $\rho$ is the liquid density, $N\geq3$ is the number of space dimensions, $\kappa$ is the polytropic exponent, $P_{0}$ is the ambient pressure of the gas in the bubble,  $R_{0}$ is the ambient radius of bubble, $\sigma$ is the surface tension. Throughout this work we use subscripts to denote derivatives.  We assume that the far--field pressure is constant. In \eqref{eq:Rayleigh_nec} we also suppose that the gas in the bubble is an ideal and obeys the polytropic law. Let us also remark that we do not consider the case of $N=2$ since in this case Eqs. \eqref{eq:Rayleigh_ec}, \eqref{eq:Rayleigh_nec} contain logarithmic terms and, therefore, are singular.

Eqs. \eqref{eq:Rayleigh_ec}, \eqref{eq:Rayleigh_nec} are $N$--dimensional generalizations of the Rayleigh equation for the empty \cite{Rayleigh1917} and gas--filled cavity \cite{Plesset1949,Plesset1977} correspondingly. Note that Eqs.\eqref{eq:Rayleigh_ec}, \eqref{eq:Rayleigh_nec} were also proposed in \cite{Prosperetti2004}.

It is also worth noting that some analytical solutions of the Rayleigh equation for the spherical bubble motion in a non-Newtonian liquid were found in \cite{Polyanin1992,Polyanin1994} in the form of a quadrature for arbitrary $p$ and $\sigma$.


We introduce the non--dimensional variables
$T=T_{c}t,\, R=R_{0}u$ in Eq.\eqref{eq:Rayleigh_ec}, where $T_{c}$ is the collapse time in $N$ dimensions given by \cite{Klotz2013}
\begin{equation}
T_{c}^{2}=\frac{\pi}{2N(N-2)}\left(\frac{\Gamma\left(\frac{1}{2}+\frac{1}{N}\right)}
{\Gamma\left(1+\frac{1}{N}\right)}\right)^{2}\frac{R_{0}^{2}\rho}{p}=
\xi^{2}\frac{R_{0}^{2}\rho}{p},
\end{equation}
and $\xi$ is the $N$--dimensional generalization of the Rayleigh factor. Using these variables from \eqref{eq:Rayleigh_ec} we get
\begin{equation}
u u_{tt}+\frac{N}{2}u_{t}^{2}=-(N-2)\xi^{2}.
\label{eq:Rayleigh_ec_non_dim}
\end{equation}

Further we consider two cases of Eq. \eqref{eq:Rayleigh_nec}. In the first case we neglect the surface tension. In the second case we investigate influence of the surface tension on the bubbles' motion.

At $\sigma=0$ we use the following non--dimensional variables $T=\omega_{0}^{-1}t,\, R=R_{0}u$  in Eq. \eqref{eq:Rayleigh_nec} to obtain
\begin{equation}
uu_{tt}+\frac{N}{2}u_{t}^{2}=\frac{1}{N\kappa}\left[u^{-N\kappa}-\beta\right],
\label{eq:gf_cavity_1}
\end{equation}
where $\omega_{0}^{2}=N(N-2)\kappa P_{0}/(\rho R_{0}^{2})$ is a $N$--dimensional generalization of bubble's natural frequency and $\beta=p/P_{0}$.

In the case of non--zero surface tension we use the non--dimensional variables $T=\widetilde{\omega}_{0}^{-1}t$, $R=R_{0}u$ in \eqref{eq:Rayleigh_nec}, where
$\widetilde{\omega}_{0}^{2}=(N-2)/(\rho R_{0}^{2})(N\widetilde{P}_{0}\kappa-(N-1)\sigma/R_{0})$.
As a result we get
\begin{equation}
u u_{tt}+\frac{N}{2}u_{t}^{2}=\alpha u^{-N\kappa}-\tilde{\beta}-S u^{-1}.
\label{eq:st_3}
\end{equation}
Here $\alpha=(N-2)\widetilde{P}_{0}/(\widetilde{\omega}_{0}^{2}\rho R_{0}^{2})$, $\tilde{\beta}=(N-2)p/(\widetilde{\omega}_{0}^{2}\rho R_{0}^{2})$, $S=(N-1)(N-2)\sigma/(\widetilde{\omega}_{0}^{2}\rho R_{0}^{3})$, $\widetilde{P}_{0}=P_{0}+(N-1)\sigma/R_{0}$.

Multiplying \eqref{eq:Rayleigh_ec_non_dim}, \eqref{eq:gf_cavity_1} and \eqref{eq:st_3} by $2u^{N-1}u_{t}$ and integrating the results with respect to $t$ we obtain
\begin{equation}
u_{t}^{2}=C_{1}u^{-N}-\frac{2(N-2)}{N}\xi^{2},
\label{eq:Rayleigh_ec_non_dim_fi}
\end{equation}
\begin{equation}
u_{t}^{2}=\frac{2}{N^{2}\kappa(1-\kappa)}u^{-N\kappa}-\frac{2\beta}{N^{2}\kappa}+C_{1}u^{-N},
\label{eq:gf_cavity_3}
\end{equation}
\begin{equation}
u_{t}^{2}=\frac{2\alpha}{N(1-\kappa)}u^{-N\kappa}-\frac{2\tilde{\beta}}{N}-\frac{2S}{N-1}u^{-1}+C_{1}u^{-N},
\label{eq:st_7}
\end{equation}
where $C_{1}$ is an integration constant. Note that we do not consider the isothermal case ( $\kappa=1$ ).

As far as $N\geq3$ and $\kappa>1$ it can be seen that $C_{1}$ has to be greater than zero for solutions of Eqs. \eqref{eq:Rayleigh_ec_non_dim_fi}, \eqref{eq:gf_cavity_3}, \eqref{eq:st_7} to be real. We can also consider Eqs. \eqref{eq:Rayleigh_ec_non_dim_fi}, \eqref{eq:gf_cavity_3}, \eqref{eq:st_7} as the energy conservation laws. Thus, the constant $C_{1}$ can be considered as the total energy and is greater than zero.

Below we construct the general solution of Eq. \eqref{eq:Rayleigh_ec_non_dim_fi} in the explicit form. We also show that the general solution of Eqs. \eqref{eq:gf_cavity_3}, \eqref{eq:st_7} in the explicit form can be constructed for certain values of the polytropic exponent. To this end we use an approach suggested in \cite{Kudryashov2014}. The main point of this approach is to transform Eqs. \eqref{eq:Rayleigh_ec_non_dim_fi}, \eqref{eq:gf_cavity_3} and \eqref{eq:st_7} into one of the equations for the Weierstrass or Jacobi elliptic functions (see, e.g., \cite{Whittaker}). It can be achieved by means of the Sundman transformation (see, e.g., \cite{Sundman1913,Meleshko,Nucci}) combined with a power--type transformation. These transformations have the form
\begin{equation}
dt=u^{\delta}d\tau,\quad u=v^{\epsilon}.
\label{eq:main_t}
\end{equation}
Here $\tau$ and $v$ are new independent and dependent variables correspondingly, $\delta$ and $\epsilon\neq0$ are real numbers. Further, we apply transformations \eqref{eq:main_t} to construct the general solutions of Eqs. \eqref{eq:Rayleigh_ec_non_dim_fi}, \eqref{eq:gf_cavity_3}.

Let us finally remark that one can express the general solution of Eq. \eqref{eq:st_7} implicitly in the form of the following quadrature
\begin{equation}
\label{eq:st_7_a}
 \pm \bigintss\frac{du}{\sqrt{\frac{2\alpha}{N(1-\kappa)}u^{-N\kappa}-\frac{2\tilde{\beta}}{N}-\frac{2S}{N-1}u^{-1}+C_{1}u^{-N}}}=t-t_{0},
\end{equation}
where $t_{0}$ is an arbitrary constant. In the same way, we can present solutions of Eqs. \eqref{eq:Rayleigh_ec_non_dim_fi} and \eqref{eq:gf_cavity_3} implicitly in the form of quadratures. However, our main goal is to find explicit expressions for general solutions of Eqs. \eqref{eq:Rayleigh_ec_non_dim_fi}, \eqref{eq:gf_cavity_3}, \eqref{eq:st_7}.

\section{Empty hyperspherical bubble}

In this section we consider the problem of the motion of the empty hyperspherical  bubble. Let us apply transformations \eqref{eq:main_t} with $\delta=N+1$ and $\epsilon=1/N$ to Eq. \eqref{eq:Rayleigh_ec_non_dim_fi}. As a result we get
\begin{equation}
v_{\tau}^{2}=-2(N-2)N\xi^{2}v^{4}+C_{1}N^{2}v^{3}.
\label{eq:emty_cavity_1}
\end{equation}
The general solution of Eq. \eqref{eq:emty_cavity_1} has the form
\begin{equation}
v=\frac{4C_{1}N}{C_{1}^{2}N^{3}(\tau-\tau_{0})^{2}+8(N-2)\xi^{2}}.
\label{eq:emty_cavity_3}
\end{equation}
Using \eqref{eq:main_t}  and \eqref{eq:emty_cavity_3} we find the general solution of \eqref{eq:Rayleigh_ec_non_dim}:
\begin{equation}
u=\left[\frac{4C_{1}N}{C_{1}^{2}N^{3}(\tau-\tau_{0})^{2}+8(N-2)\xi^{2}}\right]^{1/N}, \quad
t=\int\limits_{0}^{\tau}u^{N+1}(\zeta)d\,\zeta.
 \label{eq:emty_cavity_5}
\end{equation}
Note that throughout this work we denote by $\zeta$ and $\tau_{0}$ a dummy integration variable and an integration constant correspondingly.

\begin{figure}[!t]
\center
\includegraphics[width=0.5\textwidth]{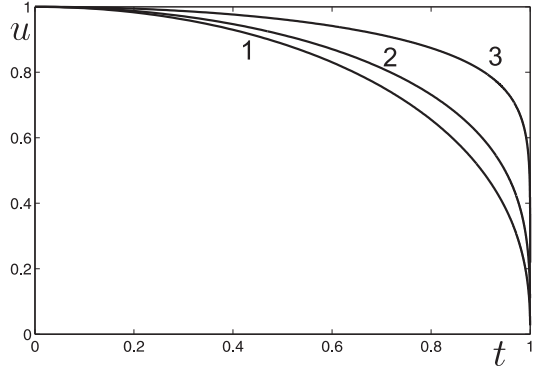}
\caption{Exact solution \eqref{eq:emty_cavity_11} of  Eq. \eqref{eq:Rayleigh_ec_non_dim} at $N=3$ (curve 1), $N=5$ (curve 2) and $N=15$ (curve 3).}
\label{f1}
\end{figure}

The integral in \eqref{eq:emty_cavity_5} can be calculated analytically and is expressed via the hypergeometric function. This expression has the form
\begin{equation}
\begin{gathered}
t=\left(\frac{C_{1}N}{2(N-2)\xi^{2}}\right)^{1+\frac{1}{N}}
\Bigg[(\tau-\tau_{0})F\left\{\frac{1}{2},1+\frac{1}{N};\frac{3}{2};-\frac{C_{1}^{2}N^{3}}{8(N-2)\xi^{2}}(\tau-\tau_{0})^{2}\right\}+\\
+\tau_{0}F\left\{\frac{1}{2},1+\frac{1}{N};\frac{3}{2};-\frac{C_{1}^{2}N^{3}}{8(N-2)\xi^{2}}\tau_{0}^{2}\right\}\Bigg].
 \label{eq:emty_cavity_5_a}
 \end{gathered}
\end{equation}
Solving the first equality from \eqref{eq:emty_cavity_5} for $\tau$ and substituting the result into \eqref{eq:emty_cavity_5_a} we find the general closed--form solution of Rayleigh equation \eqref{eq:Rayleigh_ec_non_dim} in the $N$ dimensional case
\begin{equation}
\begin{gathered}
\pm t=\left(\frac{C_{1}N}{2(N-2)\xi^{2}}\right)^{1+\frac{1}{N}}
\Bigg[\pm\tau_{0}F\left\{\frac{1}{2},1+\frac{1}{N};\frac{3}{2};-\frac{C_{1}^{2}N^{3}}{8(N-2)\xi^{2}}\tau_{0}^{2}\right\}+\\
+ \frac{2}{N\sqrt{C_{1}}}\left(\frac{1}{u^{N}}-\frac{2(N-2)\xi^{2}}{C_{1}N}\right)^{1/2}
F\left\{\frac{1}{2},1+\frac{1}{N};\frac{3}{2};\frac{C_{1}N}{2(N-2)\xi^{2}}\left[\frac{2(N-2)\xi^{2}}{C_{1}N}-\frac{1}{u^{N}}\right]\right\}\Bigg].
 \label{eq:emty_cavity_5_b}
 \end{gathered}
\end{equation}
The sing $\pm$ in the left--hand side of Eq. \eqref{eq:emty_cavity_5_b} corresponds to the invariance of Eq. \eqref{eq:Rayleigh_ec_non_dim} under the transformation $t \rightarrow -t$.   The sign $\pm$ in the right--hand side of Eq. \eqref{eq:emty_cavity_5_b} corresponds to two branches of $t$ as a function of $u$, that is for a given $u$ we have two values of $t$. From the physical point of view, the first branch (the plus sign) corresponds to the collapse motion of the bubble and the second branch (the minus sign) corresponds to the growth motion of the bubble.

Let us consider the problem of the collapse motion of the empty hyperspherical bubble. Initial conditions corresponding to this problem are the following:
\begin{equation}
u(0)=1,\quad u_{t}(0)=0.
 \label{eq:emty_cavity_7}
\end{equation}

\begin{figure}[!t]
\center
\includegraphics[width=0.5\textwidth]{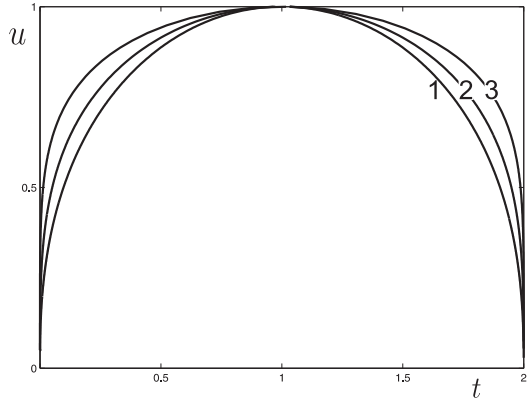}
\caption{Exact solution of Cauchy problem \eqref{eq:emty_cavity_5_a}, \eqref{eq:emty_cavity_15} at $N=3$ (curve 1), $N=5$ (curve 2) and $N=9$ (curve 3).}
\label{f2}
\end{figure}

Using \eqref{eq:emty_cavity_7} we find that $C_{1}=2(N-2)\xi^{2}/N$ and $\tau_{0}=0$. Substituting this values of $C_{1}$ and $\tau_{0}$ into \eqref{eq:emty_cavity_5_b} we get
\begin{equation}
t=\frac{1}{\xi}\left(\frac{2}{N(N-2)}\right)^{1/2}\left(\frac{1}{u^{N}}-1\right)^{1/2}
F\left\{\frac{1}{2},1+\frac{1}{N};\frac{3}{2};1-\frac{1}{u^{N}}\right\}.
 \label{eq:emty_cavity_11}
\end{equation}
We demonstrate solution \eqref{eq:emty_cavity_11} at different values of the parameter $N$ in Fig.\ref{f1}. We can see that the collapse time decreases when space dimension $N$ increases. As a consequence we see that the collapse motion of the bubble becomes more violent.

Let us consider the case of the growth and collapse motion of the empty hyper--spherical bubble. We suppose that $C_{1}$ has the same value as in the previous case and $\tau_{0}=2\sqrt{20^{N}-1}/(\sqrt{2N(N-2)}\xi)$. Taking into account these values of $C_{1}$ and $\tau_{0}$ we obtain the following initial conditions
\begin{equation}
u(0)=0.05,\quad u_{t}(0)=(N-2)\tau_{0}\xi^{2}.
 \label{eq:emty_cavity_15}
\end{equation}
We demonstrate solution \eqref{eq:emty_cavity_5_a} corresponding to initial conditions \eqref{eq:emty_cavity_15} in Fig.\ref{f2} at various values of $N$. From Fig.\ref{f2} we can see that the growth and collapse time decreases while the number of dimensions increases and bubble motion becomes more violent.  Let us also note that solutions presented in Fig.\ref{f2} correspond to both signs in Eq. \eqref{eq:emty_cavity_5_b} (that is they describe the growth and collapse motion of the bubble), and, consequently, for a given $u$ we have two values of $t$.

In this Section we have found the general analytical solution of the Rayleigh equation for the empty hyperspherical bubble. We have considered the collapse and the growth and collapse motion. Corresponding exact solutions have been obtained.

\section{Gas--filled hyperspherical bubble}

In this section we study the case of the gas--filled hyperspherical bubble. Let us note that physically possible values of the polytropic exponent for a monatomic ideal gas with $N$ translation degrees of freedom lie on the interval $1<\kappa\leq(N+2)/N$. Since $N\geq3$ we have that $1<\kappa\leq5/3$. The value of the polytropic exponent for polyatomic ideal gases is less than that for monoatomic ideal gases. Consequently, in all physically realistic cases the polytropic exponent belong to the interval $1<\kappa\leq 5/3$.

Substituting transformations \eqref{eq:main_t} into Eq.\eqref{eq:gf_cavity_3} and requiring that the resulting equation will be one of the equations for elliptic function we obtain a system of equations for the parameters $\delta$, $\epsilon$ and $\kappa$. Taking into account the above mentioned constraint on $\kappa$ and solving this system of equations we find that Eq. \eqref{eq:gf_cavity_3} can be transformed to one of the equations for the elliptic functions only at $\kappa=3/2$ and $\kappa=4/3$. Below, we consider these two values of $\kappa$ in detail.

\begin{figure}[!tp]
\center
\includegraphics[width=0.5\textwidth]{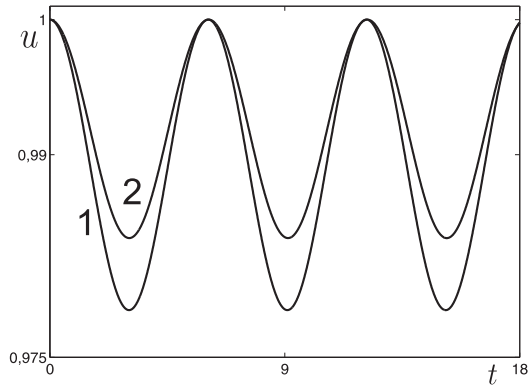}
\caption{Exact solution \eqref{eq:gf_cavity_11} of  Eq. \eqref{eq:gf_cavity_1} corresponding to initial conditions \eqref{eq:gf_cavity_11_a} at $N=3$ (curve 1) and $N=4$ (curve 2).}
\label{f_gf_1}
\end{figure}

\begin{figure}[!tp]
\center
\includegraphics[width=0.5\textwidth]{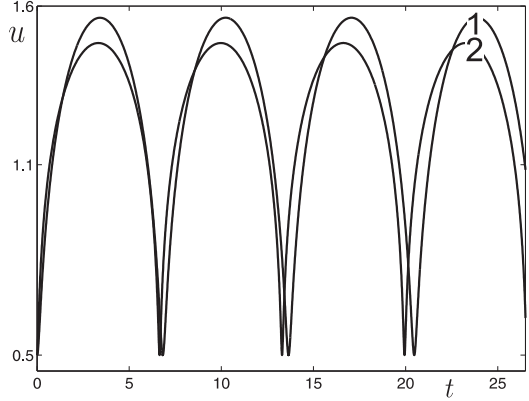}
\caption{Exact solution \eqref{eq:gf_cavity_17a} of  Eq. \eqref{eq:Rayleigh_nec} corresponding to initial conditions \eqref{eq:gf_cavity_17_b} at $N=4$ (curve 1) and $N=6$ (curve 2).}
\label{f3}
\end{figure}

Let us suppose that $\kappa=3/2$. Using transformations \eqref{eq:main_t} with $\delta=1+N/2$ and $\epsilon=-2/N$ from \eqref{eq:gf_cavity_3} we get
\begin{equation}
v_{\tau}^{2}=-\frac{2}{3}v^{3}+\frac{C_{1}N^{2}}{4}v^{2}-\frac{\beta}{3}.
\label{eq:gf_cavity_7}
\end{equation}
The general solution of Eq. \eqref{eq:gf_cavity_7} is expressed via the Weierstrass elliptic function and has the form
\begin{equation}
v=\frac{N^{2}C_{1}}{8}-\wp\left\{\frac{1}{\sqrt{6}}(\tau-\tau_{0}),g_{2},g_{3}\right\}, \quad
g_{2}=3\left(\frac{N^{2}C_{1}}{4}\right)^{2},\quad g_{3}=2\beta-\left(\frac{N^{2}C_{1}}{4}\right)^{3}.
\label{eq:gf_cavity_9}
\end{equation}
Using \eqref{eq:main_t} we find the general solution of \eqref{eq:gf_cavity_1} at $\kappa=3/2$
\begin{equation}
u=\left[\frac{N^{2}C_{1}}{8}-\wp\left\{\frac{1}{\sqrt{6}}(\tau-\tau_{0}),g_{2},g_{3}\right\}\right]^{-2/N},\quad
t=\int\limits_{0}^{\tau}u^{1+\frac{N}{2}}(\zeta)d\zeta.
\label{eq:gf_cavity_11}
\end{equation}
Note that solution \eqref{eq:gf_cavity_11} represents $u$ as a periodic function of $\tau$ and, therefore, as a periodic function of $t$. Thus, infinitely many values of $t$ correspond to a single value of $u$. The same is true for solutions of Eqs. \eqref{eq:gf_cavity_3} and \eqref{eq:st_7} which are presented below. Accordingly, we see that for a given $u$ we have several values of $t$  in plots of these solutions which are presented in Figs. \ref{f_gf_1}, \ref{f3} and \ref{f4}.

Let us study solution \eqref{eq:gf_cavity_11}. We use the following initial conditions
\begin{equation}
u(0)=1, \quad u_{t}(0)=0,
\label{eq:gf_cavity_11_a}
\end{equation}
and suppose that $\beta=1.05$. The plots of solution \eqref{eq:gf_cavity_11} at $N=3$ and $N=4$ are presented in Fig.\ref{f_gf_1}. From Fig.\ref{f_gf_1} we can see that the period of the solution changes slightly with $N$, although the amplitude of the solution changes considerably.

Now we consider the case of $\kappa=4/3$. Using transformations \eqref{eq:main_t} with $\delta=1+N/3$ and $\epsilon=-3/N$ from Eq. \eqref{eq:gf_cavity_3} we obtain the equation
\begin{equation}
v_{\tau}^{2}=-\frac{1}{2}v^{4}+\frac{C_{1}N^{2}}{9}v^{3}-\frac{\beta}{6}.
\label{eq:gf_cavity_15}
\end{equation}
The general solution of Eq. \eqref{eq:gf_cavity_15} can be expressed via one of the Jacobi elliptic functions. However it is more convenient to express this solution via the Weierstrass elliptic function. Indeed, let the parameter $\mu$ be a real solution of the equation
\begin{equation}
2187\beta\mu^{4}-16N^{8}(C_{1}\mu-1)=0.
\end{equation}

Then the general solution of \eqref{eq:gf_cavity_15} has the form
\begin{equation}
\begin{gathered}
v=\frac{324N^{2}\mu^{2}\wp\{\tau-\tau_{0},g_{2},g_{3}\}+4N^{6}(C_{1}\mu-1)}
{1458\mu^{3}\wp\{\tau-\tau_{0},g_{2},g_{3}\}-9N^{4}\mu(C_{1}\mu-2)}, \\
g_{2}=\frac{4N^{8}(C_{1}\mu-1)}{6561\mu^{4}}=\frac{\beta}{12},\quad
g_{3}=\frac{C_{1}^{2}N^{12}(C_{1}\mu-1)}{1062882\mu^{4}}=\frac{\beta C_{1}^{2} N^{4}}{7776}.
\label{eq:gf_cavity_17}
\end{gathered}
\end{equation}
Using transformations \eqref{eq:main_t} we find the general solution of \eqref{eq:gf_cavity_1} at $\kappa=4/3$
\begin{equation}
\begin{gathered}
u=\left[\frac{324N^{2}\mu^{2}\wp\{\tau-\tau_{0},g_{2},g_{3}\}+4N^{6}(C_{1}\mu-1)}
{1458\mu^{3}\wp\{\tau-\tau_{0},g_{2},g_{3}\}-9N^{4}\mu(C_{1}\mu-2)}\right]^{-3/N}, \quad
t=\int\limits_{0}^{\tau}u^{1+\frac{N}{3}}(\zeta)d\zeta.
\label{eq:gf_cavity_17a}
\end{gathered}
\end{equation}
In this case ($\kappa=4/3$) possible physical values of $N$ are 3, 4, 5 and 6. In Fig.\ref{f3} we demonstrate solution \eqref{eq:gf_cavity_17a} at $\beta=1$ and $N=4$ and $N=6$ taking into account the following initial conditions
\begin{equation}
u(0)=1/2,\quad u_{t}(0)=0.
\label{eq:gf_cavity_17_b}
\end{equation}
From Fig.\ref{f3} we see that for these initial conditions the amplitude and period of bubble oscillations decrease with increasing of space dimensions.
If we consider solutions in Fig. \ref{f3} they might seem to have sharp peaks and to be not differentiable. However, using an enlarged scale in Fig. \ref{f3}, one can find that solutions presented in this figure have smooth peaks and are differentiable functions as it follows from Eq. \eqref{eq:gf_cavity_17a}. The same is true for solutions presented in Fig. \ref{f4}.

\begin{figure}[!tp]
\center
\includegraphics[width=0.5\textwidth]{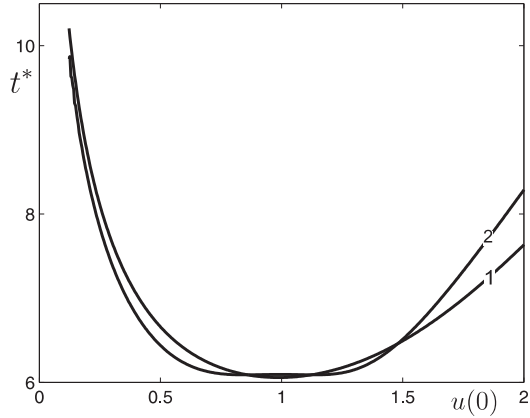}
\caption{Dependence of the solution's \eqref{eq:gf_cavity_17a} period $t^{*}$ on the initial bubble strain at $N=3$ (curve 1) and $N=6$ (curve 2).}
\label{f3a}
\end{figure}

We also present dependence of the solution's \eqref{eq:gf_cavity_17} period $t^{*}$ on the initial bubble strain, i.e. on $u(0)$, in Fig.\ref{f3a}, assuming that $u_{t}(0)=0$ and $\beta=1.05$. We can see that, if the initial strain is greater than 1, the period increases faster at $N=6$ than at $N=3$. The opposite is true for initial strain less then 1.

Finally, we remark that solutions \eqref{eq:gf_cavity_11}, \eqref{eq:gf_cavity_17a} have a real period since $\beta>0$, $C_{1}>0$ and $N\geq 3$. Consequently, only the periodic motion is possible for the gas--filled hyperspherical bubble at $\kappa=3/2$ and $\kappa=4/3$.

In this Section we have considered the Rayleigh equation for the gas--filled hyperspherical bubble. We have shown that the general solution of this equation can be constructed for certain values of the polytropic exponent. Corresponding solutions have been found and analyzed.

\section{The case of non--zero surface tension}

Let us consider the Rayleigh equation including the effect of the surface tension in the $N$--dimensional case. Applying transformations \eqref{eq:main_t} to Eq. \eqref{eq:st_7} we find that Eq. \eqref{eq:st_7} can be transformed into one of the equations for the  elliptic functions only in the case of $\kappa=4/3$ and $N=3$. Indeed, using \eqref{eq:main_t} with $\delta=2$ and $\epsilon=-1$ from Eq. \eqref{eq:st_7} at $\kappa=4/3$ and $N=3$ we get
\begin{equation}
v_{\tau}^{2}=-2\alpha v^{4}+C_{1} v^{3}-S v -\frac{2\tilde{\beta}}{3}.
\label{eq:st_9}
\end{equation}
We use the same approach as in the previous Section for finding the general solution of Eq. \eqref{eq:st_9}. Let the parameter $\mu$ be a real solution of the equation
\begin{equation}
\tilde{\beta} \mu^{4}+3\mu(S\mu^{2}-4C_{1})+48\alpha=0.
\end{equation}

\begin{figure}[!tp]
\center
\includegraphics[width=0.5\textwidth]{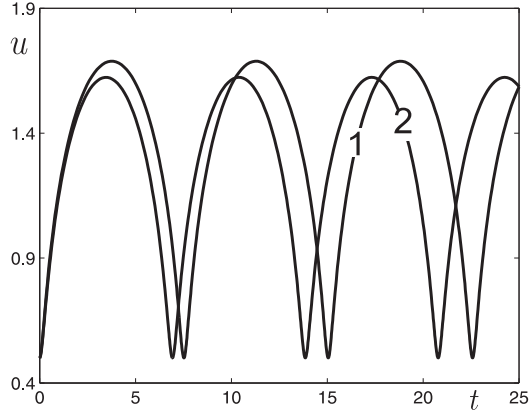}
\caption{Exact solution \eqref{eq:st_15} of  Eq. \eqref{eq:st_3} corresponding to initial conditions \eqref{eq:gf_cavity_17_b} (curve 1) and  exact solution \eqref{eq:st_15} at $\sigma=0$ corresponding to the same initial conditions (curve 2). }
\label{f4}
\end{figure}

Then the general solution of \eqref{eq:st_9} has the form
\begin{equation}
\begin{gathered}
v=\frac{8\mu^{2}\wp\{\tau-\tau_{0},g_{2},g_{3}\}+8C_{1}\mu-32\alpha-S\mu^{3}}
{4\mu^{3}\wp\{\tau-\tau_{0},g_{2},g_{3}\}-2\mu(C_{1}\mu-8\alpha)},\vspace{0.1cm}\\
g_{2}=\frac {\mu ( S\mu^{3}+64\alpha)  C_{1}-16\alpha
 ( S\mu^{3}+16\alpha)}{4{\mu}^{4}}=\frac{SC_{1}}{4}+\frac{4\alpha\beta}{3},\\
g_{3}=\frac { ( 4C_{1}-S\mu^{2} )  ( C_{1}^{2}\mu-4 \alpha C_{1}-\alpha S\mu^{2} ) }{8\mu^{4}}=
\frac{\alpha S^{2}}{8}+\frac{\beta C_{1}^{2}}{24}.
\label{eq:st_11}
\end{gathered}
\end{equation}

Using \eqref{eq:main_t} we obtain the general solution of \eqref{eq:st_3} at $\kappa=4/3$:
\begin{equation}
\begin{gathered}
u=\frac{4\mu^{3}\wp\{\tau-\tau_{0},g_{2},g_{3}\}-2\mu(C_{1}\mu-8\alpha)}
{8\mu^{2}\wp\{\tau-\tau_{0},g_{2},g_{3}\}+8C_{1}\mu-32\alpha-S\mu^{3}}, \quad
t=\int\limits_{0}^{\tau}u^{2}(\zeta)d\zeta.
\label{eq:st_15}
\end{gathered}
\end{equation}

Let us study the influence of the surface tension on the bubble motion. We consider solution \eqref{eq:st_15} corresponding to the bubble with the ambient radius $R_{0}=10^{-6}\mbox{m}$ surrounding by water with the ambient pressure $P_{0}=101.3\mbox{kPa}$. We also suppose that the far--field pressure is equal to the ambient pressure and initial conditions \eqref{eq:gf_cavity_17_b} hold. The plot of this solution is presented in Fig.\ref{f4}. We also show solution \eqref{eq:st_15} corresponding to $\sigma=0$ in Fig.\ref{f4}. From Fig.\ref{f4} we can see that the period and magnitude of the bubble motion change, while a type of the motion remains the same. Let us note that with increasing of the bubble's ambient radius the influence of the surface tension considerable decreases. We can also see that the bubble motion remains periodic in the case of non--zero surface tension as far as $\alpha>0$, $\beta>0$, $S>0$ and $C_{1}>0$.

In this Section we have found the general solution of the Rayleigh equation for the gas--filled bubble in the three--dimensional case taking into account the surface tension. We have shown that the surface tension mainly influences on the bubble motion for relatively small values of the bubble ambient radius.

\section{Conclusion}

In this work we have considered the Rayleigh equation both for the case of the empty and gas--filled hyperspherical bubble. We have constructed the general analytical solution of this equation for the empty bubble in $N$ dimensions. The collapse and growth and collapse motion of the empty hyperspherical bubble has been analyzed. We have also shown that the general solution of the Rayleigh equation for the gas--filled hyperspherical bubble can be found for certain values of the polytropic exponent. We have constructed the general solutions of the Rayleigh equation for these cases. We have discussed dependence of these solutions on space dimensions.  We have also considered influence of the surface tension on the bubble motion and constructed corresponding general solution of the Rayleigh equation in the three--dimensional case. To the best of our knowledge, our solutions have been obtained for the first time.

\section{Acknowledgments}
Authors are grateful to anonymous referees for their valuable comments and suggestions. This research was supported by Russian Science Foundation grant No. 14-11-00258.

\end{document}